\documentclass[runningheads]{llncs}
\usepackage{graphicx}
\usepackage{lineno,hyperref}
\usepackage[linesnumbered,ruled,vlined]{algorithm2e}
\usepackage{amsmath}
\usepackage{fontawesome}

\begin{document}
%
\title{Discovering Business Process Simulation Models in the Presence of Multitasking}
%
\titlerunning{Discovering BPS Models in the Presence of Multitasking}
%
\author{Bedilia Estrada-Torres\inst{1,2}\,\textsuperscript{\faEnvelopeO}\,\orcidID{0000-0001-7943-276X} \and
Manuel Camargo \inst{1}\orcidID{0000-0002-8510-1972} \and
Marlon Dumas\inst{1}\orcidID{0000-0002-9247-7476} \and
Maksym Yerokhin\inst{1}}
\authorrunning{B. Estrada-Torres et al.}
%
%

\institute{University of Tartu, Tartu, Estonia \email{\{estrada,manuel.camargo,marlon.dumas,maksym.yerokhin\}@ut.ee}\and
Universidad de Sevilla, Sevilla, Spain \\
\email{iestrada@us.es}}

\maketitle              
\begin{abstract}

Business process simulation is a versatile technique for analyzing business processes from a quantitative perspective. A well-known limitation of process simulation is that the accuracy of the simulation results is limited by the faithfulness of the process model and simulation parameters given as input to the simulator. To tackle this limitation, several authors have proposed to discover simulation models from process execution logs so that the resulting simulation models more closely match reality.
Existing techniques in this field assume that each resource in the process performs one task at a time. In reality, however, resources may engage in multitasking behavior. Traditional simulation approaches do not handle multitasking. Instead, they rely on a resource allocation approach wherein a task instance is only assigned to a resource when the resource is free. This inability to handle multitasking leads to an overestimation of execution times. This paper proposes an approach to discover multitasking in business process execution logs and to generate a simulation model that takes into account the discovered multitasking behavior. The key idea is to adjust the processing times of tasks in such a way that executing the multitasked tasks sequentially with the adjusted times is equivalent to executing them concurrently with the original processing times. The proposed approach is evaluated using a real-life dataset and synthetic datasets with different levels of multitasking. The results show that, in the presence of multitasking, the approach improves the accuracy of simulation models discovered from execution logs.

\end{abstract}

%
%
%


\section{Introduction}
\label{sec:introduction}

Business process simulation (BPS) is a widely used  technique for analyzing quantitative properties of business processes. The basic idea of BPS is to execute a large number of instances of a process, based on a process model and a number of simulation parameters, in order to collect performance measures such as waiting times of tasks, processing times, execution cost, and cycle time~\cite{vanderAalst_2010_BPSRevisited,Dumas_2018_FBPM}. BPS tools (simulators) allow analysts to identify performance bottlenecks~\cite{Rusinaite_2016_SharedResource} and to estimate how a given change to a process may affect its performance~\cite{Sander_2018_QEffects_Simulation}. 

The accuracy of a process simulation, and hence the usefulness of the conclusions drawn from it, is to a large extent dependent on how faithfully the process model and simulation parameters capture the observed reality. Traditionally, process models are manually designed by analysts for the purpose of communication and documentation. As such, these models do not capture all the intricacies of how the process is actually performed. In particular, manually designed process models tend to focus on frequent pathways, leaving aside exceptions. Yet, in many cases, exceptions occur in a non-negligible percentage of instances of a process. Moreover, simulation parameters for BPS are traditionally estimated based on expert intuition, sampling, and manual curve fitting, which do not always lead to an accurate reflection of reality~\cite{Rusinaite_2016_SharedResource}. 

To tackle these limitations, several authors have advocated the idea of automatically discovering simulation models from business process execution logs (also known as \emph{event logs})~\cite{MartinDC16,Camargo_2019_SIMODCoRR}. 
Simulation models discovered in this way are generally more faithful since they capture not only common pathways, but also exceptional behavior. Moreover, automated approaches to simulation model discovery typically explore a larger space of options when tuning the simulation parameters compared to what an analyst is able to explore manually.

The automated discovery of BPS models from event logs opens up the possibility of capturing resource behavior at a finer granularity than manual BPS modeling approaches. In particular, Martin et al.~\cite{MartinDCS20} demonstrated the possibility of discovering fine-grained resource availability timetables from event logs and the benefits of using these timetables to enhance the accuracy of BPS models.

Inspired by this possibility, this paper studies the problem of discovering another type of resource behavior, namely multitasking, from an event log. Multitasking refers to the situation where a resource executes multiple task instances simultaneously, meaning that the resource divides its attention across multiple active task instances~\cite{Ouyang_2010_ResourceRequirements}. The inability to capture multitasking behavior has been identified as a limitation of existing BPS approaches, for example in~\cite{vanderAalst_2015_SimGuide}.

Concretely, the paper proposes an approach to discover multitasking behavior from an event log and to generate a BPS model that takes into account the discovered multitasking behavior.
The key idea is to adjust the processing times of task instances in such a way that executing the multitasked task instances sequentially with the adjusted times is equivalent to executing them concurrently with the original processing times. 
Once the event log is adjusted is this way, we discover a BPS model using existing BPS model discovery techniques, namely those embedded in the SIMOD tool~\cite{Camargo_2019_SIMOD}. 
The proposed approach is evaluated using a real-life dataset and synthetic datasets with different levels of multitasking. 

The rest of this article is structured as follows. Section~\ref{sec:motivation} motivates our research. 
Section~\ref{sec:background_relatedwork} introduces basic concepts and related work. 
Section~\ref{sec:formalization} describes the proposed approach. 
Finally, Section~\ref{sec:evaluation} reports on the evaluation of the approach while Section~\ref{sec:conclusions_futurework} draws conclusions and outlines directions for future work.

\section{Motivation}
\label{sec:motivation}

During business process execution, certain events are recorded which capture, for example, the moment when a task instance  started and ended, the resource that executed the task instance, etc. Such events are stored in \textit{event logs}, which can be used to analyze the performance of the process or to discover process models that faitfully reflect the actual execution of the process.

Sometimes, the events associated to a given resource in an event log may show that the resource started a task instance before completing a previous one. Hence, during some period of time, the resource performs multiple task instances simultaneously, a situation known as \emph{multitasking}.  
Multitasking arises, for example, when a resource postpones the completion of a task due to  missing information. While this information becomes available, the resource may start another task instance to avoid idle times. 

\begin{figure}
  \centering
    \includegraphics[width=\textwidth]{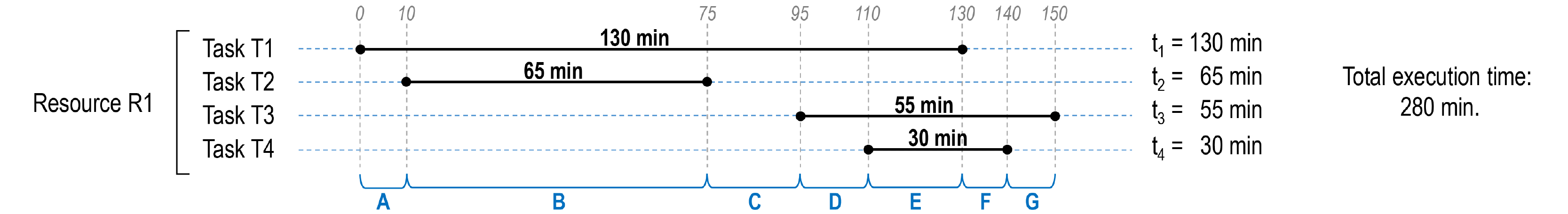}
    \caption{Example of multitasking for the resource $R1$}
    \label{fig:example_multitasking_resource_1}
\end{figure}

Figure~\ref{fig:example_multitasking_resource_1} represents a subset of four tasks carried out by resource $R1$, where each continuous line represents the duration of each task. 
These four tasks result in seven execution intervals. In intervals \textit{A}(0-10), \textit{C}(75-95) and \textit{G}(140-150) only one task is executed, T1, T1 and T3, respectively. Other segments reflect multitask execution: in \textit{B}(10-75), \textit{D}(95-110) and \textit{F}(130-140) two tasks are executed (T1, T2), (T1, T3) and (T3, T4), respectively; and in interval \textit{E}(110-130) multitasking is performed between three tasks (T1, T3, T4). These tasks may belong to one or more traces. A \textit{trace} contains the ordered sequence of events observed for a given process instance~\cite{Dumas_2018_FBPM}. An event log is composed of one or~more traces.

Given the data of this event log segment, a traditional simulator would calculate a total execution time of 280 minutes, because it would take each duration individually, one task after the other. However, in Figure~\ref{fig:example_multitasking_resource_1}, it is possible to see that all tasks are executed between the $0$ and the $150$ minute. This means that during certain intervals, the resource $R1$ divided its time and attention into more than one task. 
Therefore, it would not be correct to consider as total task execution time the time between the start and end record of the task, but the time should be distributed among all the tasks that overlap in a given period. 

Since there is usually no detailed record of the specific time that each resource spends on the execution of each task in multitasking scenarios, we consider it necessary to propose a mechanism to adjust processing times to reflect the time spent by each resource more accurately.

\section{Background and Related Work}
\label{sec:background_relatedwork}

In a simulation scenario, a \textit{work item} is created during a process simulation, when a task is ready to be executed, which can be seen as an instance of a task that will be executed~\cite{Russell_2005_WorkflowPatterns}.
A simulator tries to assign each \textit{work item} to a resource. Once that it is been assigned, the simulator determines the \textit{work item} duration and that \textit{work item} is placed in suspend mode during the assigned duration time. When the duration time ends, the \textit{work item} is considered completed and the resource is again available to be used by another \textit{work item}~\cite{Dumas_2018_FBPM}. 

Many efforts have been made to try to simulate process models as close to reality as possible. However, simplifications of these processes are still needed due to the technical limitations of certain simulators~\cite{vanderAalst_2008_BPSRight}. One of the areas of interest that still has deficiencies is related to the behavior of resources involved in a process execution. 
Several workflow resource patterns are presented in~\cite{Russell_2005_WorkflowPatterns}, describing how resources are represented and used in a workflow. 
However, behaviors such as described in \cite{vanderAalst_2010_BPSRevisited} and \cite{vanderAalst_2015_SimGuide} have not yet been fully incorporated into simulation techniques. 
On the basis of patterns described in~\cite{Russell_2005_WorkflowPatterns}, Afifi et al. presented in \cite{Afifi_2018_WRPMetamodel} and \cite{Afifi_2018_ExtensioonBPSim}, the extension of BPSim, the Business Process Simulation Standard~\cite{WfWC_2016_BPSim}. 
BPSim provides a metamodel and an electronic file format to define process models including simulation-specific parameters. 
However, a tool to support for this proposal is suggested as future work.

Ling et al.~\cite{Ling_2014_BPSResource} propose a prototype simulation tool that considers differences between resources based on their experience and on personnel movements such as recruitment, transfer and resignation. However, this proposal does not consider the possibility of performing more than one task in a given time instant. 
Although Ouyang et al.~\cite{Ouyang_2010_ResourceRequirements} point out that real business processes are resource-intensive, where multitasking situations are typical, their study focuses on proposing a conceptual model, 
in which its possible to model and schedule the use of shared material resources, such as surgical material that is shared by several doctors during a surgical operation; but unlike our proposal, the authors do not analyze real execution data, nor do they simulate proposals for the use of shared resources, since their implementation is proposed as future work.

In the approach proposed by Rusinaite et al. in~\cite{Rusinaite_2016_SharedResource}, resources (human or not) are classified as \textit{shareable resource}, to indicate a resource can be used by several activities simultaneously; and \textit{non-shareable resource}, when a resource is allocated to only one activity at a time. For the modeling and simulation a resource is defined by means of attributes of \textit{capacity} (reusable or consumable) and \textit{shareability} (shareable and non-shareable). A simulation engine was used to validate the proposal, where authors found that when two or more resources are available, average time decreased considerably as shared resources are used. From the shared use of five resources, the difference in time was less noticeable. Authors do not specify how shared times are defined and calculated. One inconvenience of this proposal is the need to know beforehand the characteristics of the resource being used. On the contrary, in our proposal, we use event logs to identify if a (human) resource has been running simultaneous tasks and to determine the fragments of time in which the tasks were executed simultaneously.

\section{Approach}
\label{sec:formalization}

\begin{figure}
  \begin{center}
    \includegraphics[width=\textwidth]{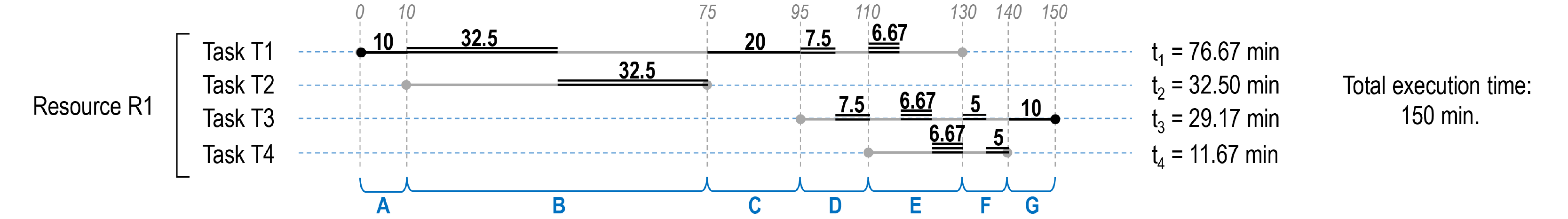}
    \caption{Example of time adjustments derived from multitasking}
    \label{fig:example_multitasking_intersections_1}
  \end{center}
\end{figure}

As explained in Section~\ref{sec:motivation}, a resource can start a \textit{work item} before finishing one or more \textit{work item} he/she started before, but simulators are not capable of taking this behavior into account. 
To cope with this lack, we propose to pre-process event logs to adjust the processing times, which is to proportionally divide the interval of execution time where different tasks intersect by the number of tasks involved. In this way, multitasking can be approached without modifying the structure and operations of the simulators.   Figure~\ref{fig:example_multitasking_intersections_1} shows how the duration of multitasked intervals (Figure~\ref{fig:example_multitasking_resource_1}) are distributed proportionally among the number of tasks in each interval. For example, in interval \textit{B}, the total time (65 minutes) is divided proportionally between tasks T1 and T2 (32.5 minutes for each); or in segment \textit{C}, three tasks are executed, so the 20 minutes of its duration are divided between tasks T1, T3 and T4. In this way, the new task execution times are more similar to the real dedication of the resource.

The objective of pre-processing event log is, on the one hand, to identify the resources that perform multitasking, determine in which time periods the multitasking execution is performed and to make an adjustment of the \textit{work item} duration times according to the multitasking periods. And, on the other hand, to determine how the multitasking execution intervals influence the general performance of the business process.   
We assume that resources are involved in only one business process at a time.

The following definitions describe step-by-step how the event log is pre-processed. To do this, we begin by formally defining the concepts of \textit{event}, \textit{trace}, \textit{event log} and \textit{work item}.

\begin{definition}[Events, Attribute] 
Let $\mathcal{E}$ be the set of all possible events 
that occur during a process execution.  
Let's assume an event $e$ can be described by means of a set of attributes $att$, where $att = \{id,type,r,st,et\}$, $id$ is the identifier of the event;
$type$ represents the event type, the activity name;
$r$ represents the resource that performs the event;
$st$ indicates the event start timestamp;
$et$ indicates the event end timestamp. 
In such a way that, for example, $att_{r}(e) = e_{r} = r_1$, where $r_1$ is a particular resource performing $e$. 
\end{definition}

\begin{definition}[Trace] 
Let $\mathcal{T}$ be the set of all possible traces defined as a sequence of events, such that, $\sigma \in \mathcal{T}, \sigma = <e_1, e_2, ..., e_n>$
\end{definition}

\begin{definition}[Event Log] 
An event log can be defined as a set of traces, $\mathcal{L} \subseteq \mathcal{T}$, where $\mathcal{L} = <\sigma_1, \sigma_2,...,\sigma_n>$
\end{definition}

\begin{definition}[Work Item] 
Let $wi$ be a work item representing an event in a process simulation, in such a way that $wi \approx e$. Therefore, a trace can be represented as a sequence of work items, such that $\sigma \in \mathcal{T}, \sigma = <wi_1, wi_2, ..., wi_n>$.

As with events, a work item has the set of attributes $att$. For example, $att_{r}(wi) = wi_{r} = r_1$, where $r_1$ is the resource that has the $wi$ assigned to it. 
\end{definition}

Multitasking can be generated by \textit{work items} generated in a single trace or by \textit{work items} belonging to different traces. In this proposal, the broadest case is considered, so all traces in which each resource participates is considered.
In order to identify the task (and \textit{work items}) in which a resource perform multitasking, the log $\mathcal{L}$ is divided into as many \textit{Segment per Resource} as there are resources in log. Each segment consists of all the \textit{work items} of each resource in $\mathcal{L}$, which will be ordered according to the start timestamp of each \textit{work item} ($wi_{st}$). 
Figure~\ref{fig:example_multitasking_resource_1} represents one \textit{Segment per Resource} ($sr_1$) with four \textit{work items} for resource $R1$.

\begin{definition}[Segment per Resource]\label{def:segment_per_resource} 
Given an event log $\mathcal{L}$, $\mathcal{R}$ represents the set of all possible resources that execute at least one work item in any trace in a log $\mathcal{L}$. Such that, 
$\forall r \in \mathcal{R}, \exists wi \subseteq \mathcal{T} \subseteq \mathcal{L} \mid wi_r = r$

Then, let $\mathcal{S}$ be the set of all possible ordered subsets of work items conforming the traces of a log, in such a way that  $\mathcal{L} = \{ sr_1, sr_2, \dots, sr_n\}$, where $\forall sr_i \in \mathcal{S}, sr_i = <wi_j,...,wi_m> \mid (wi_{j_{r}} = wi_{j+1_{r}} = \dots = wi_{m_{r}}) \wedge (wi_{j_{st}} \leq wi_{j+1_{st}} \leq \dots \leq wi_{m_{st}})$, where $1 \leq j \leq n$.

\end{definition}

Having divided $\mathcal{L}$ into different ($sr_i$), the \textit{Sweep Line algorithm}~\cite{Arge_1998_SweepLine} is applied to each $sr_i$ to identify intersection points between \textit{work items} determined by their start and end timestamps. 
For each pair of intersection points between the different \textit{work items}, \textit{auxiliary work items} ($wiaux$) are created. 

To identify the set of $wiaux$, first, for each \textit{segment per resource} $sr_i$ an ordered list of time ($ordtimes$) is created, where $ordtimes = $ \{ $point_1, point_2, ..., point_n$ \}. Each element of the list, called $points$, is a tuple $point_i = (tstamp_i, wiid_{i}, symbol_i)$, where $tstam_i$ could be a start timestamp or an end timestamp of any of the work items in $sr_i$; $wi_{i_{id}}$ is the identifier of the work item with start or end timestamp equals to $tstamp_i$; $symbol_i$ could be `+' if $tstamp_i$ corresponds to a start timestamp, or `-' if it is an end timestamp; and $wi_i$ is the complete \textit{work item} used to obtain the other values of the tuple. 

\begin{definition}[Ordered List of Times]\label{def:orderedlisttimes}
$\forall sr_i \subseteq \mathcal{L}, \exists times_i, ordtimes_i \mid \{ (wi_{j_{st}},$ $wi_{j_{id}},`+$'$), (wi_{j_{et}}, wi_{j_{id}},`-$'$), \dots, (wi_{n_{st}}, wi_{n_{id}},`+$'$), (wi_{n_{et}}, wi_{n_{id}},`-$'$)\} \wedge ordtimes_i = \{ (tstamp_k, wiid_k, symbol_k), (tstamp_{k+1}, wiid_{k+1}, symbol_{k+1}), \dots, (tstamp_l, wiid_l,$ $symbol_l) \} \wedge tstamp_k \leq tstamp_{k+1} \leq \dots \leq tstamp_l \wedge \vert times_i\vert = \vert ordtimes_i\vert$, where ($tstamp_x = wi_{x_{st}} \vee tstamp_x = wi_{x_{et}}); wiid_x = wi_{x_{id}}; symbol_x \subset \{`+$',$`-$'\}.
\end{definition}

\SetKwInput{KwInput}{Input}                
\SetKwInput{KwOutput}{Output}              

\begin{algorithm}
\DontPrintSemicolon
\scriptsize  
  \KwInput{Ordered list of times $ordtimes_i$}
  \KwOutput{List of auxiliar work items $lwiaux$}
  temp\_ids = []; 
  intervals = []; 
  lwiaux = [];
    id = 1
  
  \For{i in range(0,len(ordtimes)-1)}{
    \If{(exists(ordtimes[i+1]))}{
        \If{ordtimes[i][`symbol'] == `+')}{
            temp\_ids.append(ordtimes[i][`wiid'])
        }
        \Else{
            temp\_ids.remove(ordtimes[i][`wiid'])
        }
        intervals.append(ordtimes[i][`tstamp'], ordtimes[i+1][`tstamp'], temp\_ids)
    }
  }
  
    \For{interval in intervals}{
     \For {wiid in interval[`list\_wiid']}
        {lwiaux.append(id, interval[`start\_int'], interval[`end\_int'], interval[list\_wiid][`wiid'])
        
        id += 1
        }
  }

\caption{Creating $wiaux$ elements in a $lwiaux$}
\label{alg:lwiaux}
\end{algorithm}

Once the $ordtimes_i$ has been created, concrete intervals of time $intervals_i$ are specified, identifying also the work items $wi_n$ that are being executed for each interval, $intervals_i = \{ (start\_int_1, end\_int_1, list\_wiids_1),..., (start\_int_k, end\_int_k,$ $list_wiids_k)\}$, where $start\_int$ and $end\_int$ represent the start and end timestamp of the intersected work items collected in $list\_wiids$. 
For each element in $list\_wiids$ an auxiliar work item $wiaux$ is created, in such a way that $wiaux = (start\_int, end\_int, id, duration)$, where $wi = \{ wiaux_1,\dots, wiaux_n \}$ and $wi_{et} - wi_{st} = \sum_{n=1}^{n} wiaux_{n_d}$. 
The $duration$ of each $wiaux$ is determined by the number of $wiaux$ generated from a given $interval, duration = (end\_int - start\_int)/$ $len(list_wiids)$. For example, from $interval = (10,75,`wi_1,wi_2$'$)$ two $wiaux$ are generated $wiaux_1 = (10,75,`wi_1$'$,32.5)$, $wiaux_2 = (10,75,`wi_2$'$,32.5)$. The list $lwiaux$ contains all $wiaux$ generated.

Based on the above definitions, Algorithm~\ref{alg:lwiaux} describes how the adjustment of task execution times is performed taking into account the number of tasks that are simultaneously executed by a resource, by means of the creation of the $lwiaux$ list. 
Applying the Definition~\ref{def:orderedlisttimes} and the Algorithm \ref{alg:lwiaux} to the scenario depicted in Figures \ref{fig:example_multitasking_resource_1} and \ref{fig:example_multitasking_intersections_1}, the set of values presented in Table~\ref{tab:proposal_values} are obtained.

\begin{table}
\centering
\caption{Intermediate values obtained from Definition~\ref{def:orderedlisttimes} and Algorithm \ref{alg:lwiaux}}
\label{tab:proposal_values}
\scriptsize
\begin{tabular}{llll}
\hline

ordtimes & = 
			& \begin{tabular}[c]{@{}l@{}} \{(0, A, ‘+’), (10, B, ‘+’), (75, B, ‘-’), (95, C, ‘+’), (110, D, ‘+’), (130, A, ‘-’), \\(140, D, ‘-’), (150, C, ‘-’)\} \end{tabular} 
      & \begin{tabular}[c]{@{}l@{}} \end{tabular} \\
      
intervals & = 
			& \begin{tabular}[c]{@{}l@{}} \{(0, 10, ‘A’), (10, 75, ‘A,B’), (75, 95, ‘A’), (95, 110, ‘A,C’), (110, 130, ‘A,C,D’), \\(130, 140, ‘C,D’), (140, 150, ‘C’)\} \end{tabular} 
      & \begin{tabular}[c]{@{}l@{}} \end{tabular} \\
      
lwuiaux & = 
			& \begin{tabular}[c]{@{}l@{}} \{(0, 10, ‘A’, 10), (10, 75, ‘A’, 32.5), (10, 75, ‘B’, 32.5), (75, 95, ‘A’, 20), \\(95, 110, ‘A’, 7.5), (95, 110, ‘C’, 7.5), (110, 130, ‘A’, 6.67), (110, 130, ‘C’, 6.67), \\(110, 130, ‘D’, 6.67), (130, 140, ‘C’, 5), (130, 140, ‘D’, 5), (140, 150, ‘C’, 10)\} \end{tabular} 
      & \begin{tabular}[c]{@{}l@{}} \end{tabular} \\

\hline
\end{tabular}
\end{table} 

Given an event log $\mathcal{L}, len(\mathcal{L})$ indicates the number of work items in $\mathcal{L}$. 
And according to the above definitions it is possible to state that $lwiaux = \mathcal{L}'$, where $\mathcal{L}'$ is defined as:

\begin{definition}[Auxiliar Event Log ($\mathcal{L}'$)]
Given a $\mathcal{L}$, 
$\forall \mathcal{L} = <wi_1, wi_2, ..., wi_n>, \exists \mathcal{L}' \mid \mathcal{L} \equiv \mathcal{L}' \wedge \mathcal{L}' = <wiaux_1, wiaux_2,...,wiaux_m>$, where $wi_i = <wiaux_j,\dots,$ $wiaux_k>, 1 \leq i \leq n, 1 \leq j \leq k, m \geq len(\mathcal{L})$.
\end{definition}

From $\mathcal{L}'$ it is possible to generate a ``coalescing log" $\mathcal{L}''$ that contains a set of coalesing work items $wicoal$. Each $wicoal$ is the result of the sum of the pre-processed times ($wiaux$) of each original $wi$ in $\mathcal{L}$. 

\begin{definition}[Coalescing Log ($\mathcal{L}''$)]
$\forall \mathcal{L}=<wi_1,...,wi_n> ,\mathcal{L}'=<wiaux_1,$  $...,wiaux_m> \exists \mathcal{L}'' = <wicoal_1,...,wicoal_n> \mid 
wicoal_{i_{id}} = wi_{i_{id}} \wedge
wicoal_{i_{type}} = wi_{i_{type}} \wedge
wicoal_{i_{r}} = wi_{i_{r}} \wedge
wicoal_{i_{st}} = wi_{i_{st}} \wedge  
wicoal_{i_{et}} = (wicoal_{i_{st}} + sum_{t=1}^{m} wiaux_{t_{d}} ) \wedge
len(\mathcal{L}) = len(\mathcal{L}'') \wedge [sum_{t=1}^n(wi_{t_{et}} - wi_{t_{st}}) = sum_{t=1}^n(wicoal_{t_{et}} - wicoal_{t_{st}})]$ 
 
\end{definition}

From the above definitions we can deduce that:
$\forall \mathcal{L} \exists \mathcal{L}', \mathcal{L}'' \mid \mathcal{L} \equiv \mathcal{L}' \equiv \mathcal{L}'' \wedge len(\mathcal{L}) \leq len(\mathcal{L}') \wedge len(\mathcal{L}') \geq len(\mathcal{L}'') \wedge len(\mathcal{L}) = len(\mathcal{L}'')$

In addition, if $len(\mathcal{L}) == len(\mathcal{L}')$ there is no multitasking, because the execution times do not intersect for any work item of any resource in the event log $\mathcal{L}$ and $\forall wi_i \in \mathcal{L} \mid wi_i = \{wiaux_i\}$

The level of multitasking in a given log, is determined by the amount of overlap between the execution times of pairs of events in a log, for a given resource, in proportion to the number of total pairs of events that can be formed between the work items of each $sr_i$. 
In order to determine the level of multitasking in a given log $\mathcal{L}$, we propose a measure called \textit{\textbf{Multitasking Log Index} (\textit{MTLI}}). 
To calculate the $MTLI$ of a log $\mathcal{L}$, we based on the idea that a log is divided by grouping all the work items of a given resource ($r$), generating $sr_i \in \mathcal{S}$ (See Definition~\ref{def:segment_per_resource}). 
The multitasking of a log is derived from the overlap between the execution times of two work items executed by the same resource. Therefore, for each $sr_i$, let $WI_{sr}$ be the set of all possible work items in $sr_i$ and $SRWI_r$ be the set of all possible pairs of work items in $sr_i$.

$$SRWI_r = \{ (wi_1, wi_2) \in WI_{sr} \times WI_{sr} \mid wi_1 \neq wi_2 \wedge wi_1.r = wi_2.r \}$$

For each pair of events $(wi_1,wi_2)_i \in SRWI_r$, $1 < i < \mid SRWI_r \mid$, an overlap function is calculated as the maximum between the zero and the difference of the minimum of the end timestamps of the work items and the maximum of their start timestamps; divided by the maximum value of the duration of the two work items. 

$$ overlap(wi_1,wi_2)_i= \frac{max((min(wi_1.et, wi_2.et)-max(wi_1.st, wi_2.st)), 0)}{max((wi_1.et-wi_1.st),(wi_2.et-wi_2.st))} $$

With the previos information it is possible to calculate the \textit{Multitasking Resource Index ($MTRI_r$)}, as the index of multitasking for each $sr_i$ in the log. For each $sr_i$, all overlap values are summed; and that sum is multiplied by the value of 1 divided number of pair of events in $SRWI_r$. 

$$MTRI_r = \frac{1}{\mid SRWI_r \mid} \displaystyle\sum_{(wi_1,wi_2)_i \in SRWI_r}^{\mid SRWI_r \mid } overlap(wi_1, wi_2)_i $$

Finally, $MTLI$ is calculated as the average of all $MTRI_r$ in the log. 

$$MTLI = \frac{ \displaystyle\sum_{j = 1}^{\mid \mathcal{S} \mid} MTRI_j }{\mid \mathcal{S} \mid}, \mathcal{S} = \{ sr_1, \dots, sr_n \} $$

The \textbf{\textit{Multitasking Work Items Index (MTWII)} } is another measure related to multitasking that is calculated in a very similar way that $MTLI$, but in this case, only overlapped pairs of events are considered. The set of all possible overlapped pairs of events for a resource is defined as follows.

\begin{equation*}
\begin{split}
RWI_{o_{r}} = {} & \{ (wi_1, wi_2) \in WI_{sr} \times WI_{sr} \mid wi_1 \neq wi_2 \wedge wi_1.r = wi_2.r \\
                 & \wedge (min(wi_1.et, wi_2.et)-max(wi_1.st, wi_2.st)) > 0 \}
\end{split}
\end{equation*}

The function $overlap(wi_1, wi_2)$ is calculated the same way.  
Now, the value of $MTRI_{r}$ is calculated only for those pairs of events ovelapped~($MTRI_{o_r}$).  

$$MTRI_{o_r} = \frac{1}{\mid RWI_{o_r} \mid} \displaystyle\sum_{(wi_1,wi_2)_i \in RWI_{o_r}}^{\mid RWI_{o_r} \mid } overlap(wi_1, wi_2)_i $$

Finally, $MTWII$ is calculated as the average of all $MTRI_{o_r}$, where ${S}_{o}$ represents the set of all resources that have at least on pair of work items with multitasking.
$\mathcal{S}_{o} = \{ sr_{o_1}, \dots, sr_{o_j} \}$, where $1<i<j$; 
$sr_{o_i} = \{ wi_1, \dots, wi_k \} \mid \exists (wi_n, wi_m) \in MTRI_{o_r} $, where $1 < n,m < j; wi_n \neq wi_m$.

$$MTWII = \frac{ \displaystyle\sum_{r = 1}^{\mid \mathcal{S}_{o} \mid} MTRI_{o_r} }{\mid \mathcal{S}_{o} \mid}$$

\section{Evaluation}
\label{sec:evaluation}

The pre-processing of an event log for the identification of multitasking \textit{work items}, the overlapping time periods, the adjustment of the execution times for these \textit{work items} and the calculation of multitasking indexes is done by means of a \textit{Sweeper} Python script.  
It receives as input a base event log ($\mathcal{L}$) in eXtensible Event Stream (XES) format and generates as output an event log with the adjusted times according to the multitasking previously identified ($\mathcal{L''}$). 
The $\mathcal{L}$ must contain \textit{work items} with at least the task name, the resources that executed the \textit{work item}, and the start and end timestamps for each \textit{work item}. An identifier for each \textit{work item} is assigned during pre-processing. 
In addition, the events in the log must reflect multitasking in order to perform the analysis. 
Based on these restrictions, the evaluation was twofold and was performed using a real event log and a set of synthetic logs. Event logs\footnote{\url{https://github.com/AdaptiveBProcess/Simod/tree/master/inputs/multitasking\_logs}} and scripts\footnote{\url{https://github.com/AdaptiveBProcess/Simod/tree/master/support\_modules/multitasking}} are available online. 
The experiments were carried out on a computer using Windows 10 Enterprise (64-bit), a processor Inter Core i5-6200U, CPU 2.3GHz and 16.0 GB RAM.

In both real and synthetic cases, after generating the event logs with the adjusted times derived from multitasking, the SIMOD tool~\cite{Camargo_2019_SIMOD} was used to discover business process simulation models. This tool uses the hyper-parameter optimization technique ``to search in the space of possible configurations in order to maximize the similarity between the behavior of the simulation model and the behavior observed in the log". 
Process models are discovered using the Split Miner algorithm\cite{Augusto_2017_SplitMiner}, which considers different levels of sensibility and depends on two parameters: the parallelism threshold, epsilon($\epsilon$) that determines the quantity
of concurrent relations between events to be captured; and the percentile for frequency threshold, eta ($\eta$), that acts as a filter over the incoming and outgoing edges of each node and retains only the most frequent percentiles. Both parameters are defined in a range between 0 and 1. 
The resulting simulation models can be executed using Scylla\cite{Pufahl_2018_Scylla} and BIMP\cite{Madis_2011_BIMP}. As in \cite{Camargo_2019_SIMODCoRR}, we use BIMP because it allows a wider set of distribution probabilities to be used, thus widening the space for configuration options. 
During SIMOD executions, an objective evaluation of the results is made by means of similarity measures which will be described in more detail in the following subsections.


\subsection{Evaluation based on a Real-life Event Log}
\label{subsec:evaluation_real}

The objective of this section is to identify the actual accuracy gains of the proposal, using a real-life event log. 
The hypothesis in this scenario is that adjusting execution times derived from multitasking provides more accurate execution results, reduces the total execution time of tasks and processes; avoid over-utilization of resources due to sequential simulation of task execution; and maintains the correct alignment of the model generated according to the original model derived from the log. 
The real event log represents an academic credentials recognition (\textit{ACR log}) process in an University during the first semester of 2016. This log has 954 traces, 18 tasks, 6870 events and involves 561 resources. 


\subsubsection{Experimental setup.}
\label{subsub:reallog_setup}

The validation process is divided into the following steps: 

\begin{enumerate}

\item \textit{Create the adjusted log}. Execute the \textit{sweeper.py} script using the \textit{ACR log} to generate a new event log with the adjusted times (\textit{ACR adjusted log}). 

\item \textit{Calculate measures}. 

\begin{itemize}
\item The execution of the \textit{Sweeper} script also provides a set of values and indexes to identify the level of multitasking in the \textit{ACR log}.
\item The hyper-parameter optimization of SIMOD was used with both logs to obtain similarity measures in each case. 50 BPS models were generated using different setup combinations of processing parameters. Parameters $\epsilon$ and $\eta$ varied from 0.0 to 1.0. Each BPS model was executed 5 times, for that, 250 simulations were evaluated for each of the both event logs. 
\item Finally, Apromore\footnote{http://apromore.cs.ut.ee/} can be used for the comparison of processing times and BIMP\footnote{http://bimp.cs.ut.ee/} to analyze resource utilization values. 
\end{itemize}

\item \textit{Analyze the results}. Compare values between two logs.

\end{enumerate}


\subsubsection{Analysis of results.}
\label{subsub:reallog_analisis_results}

Executing SIMOD using the \textit{ACR log}, very similar results were obtained to those presented in \cite{Camargo_2019_SIMODCoRR}, using half number of simulations. In that proposal, the similarity measure Timed String Distance (TSD) is calculated. TSD is a modification of the distance measure called Demerau-Levinstein (DL) that assesses the similarity between two process traces. TSD allows to include a penalty related to the time difference in processing and waiting times providing a single measure of accuracy. 
In~\cite{Camargo_2019_SIMODCoRR}, TSD is equal to \textit{0.9167}. In our experiment, TSD is equal to \textit{0.906} with $\epsilon$=0.615 and $\eta$=0.559. 
Executing \textit{ACR adjusted log}, similarity measure is equal to \textit{0.929} with  $\epsilon$=0.484 and $\eta$=0.591. 

The difference between similarity values of both logs is \textit{2.54\%}. Although this value may seem low, it should not be seen by itself; it should be analyzed in relation to the amount of multitasking identified in the log. 
Table~\ref{tab:differences_in_logs} shows the characteristics related to the content of the event log.  
Out of the 18 tasks in the event log, 17 are overlapped in at least one instance (work item) within the log. From the 6870 events (\textit{work items}), 1267 are overlap with at least one other event. Out of the 561 resources involved in the log, 76 executed at least one event with multitasking. Finally, after grouping all events according to the resource that executed them, 1116776 pairs of events were identified. Of all of them, 1036 are overlapped in some period of their execution time. This last feature is very significant as it indicates that the log used for this analysis actually has a low amount of multitasked events.   
This is a possible reason why the percentage of similarity improvement was quite low (2.54\%). 
In addition, from the BPS model results generated by SIMOD, it is possible to extract the \textit{Average Cycle Time} of each simulation. When comparing the results of both logs, an improvement of approximate 14\% of the \textit{ACR adjusted log} with respect to \textit{ACR log} was obtained.

\begin{table}
\centering
\caption{Differences between original log features and features reflecting multitasking}\label{tab:differences_in_logs}
\begin{tabular}{|l|c|c|c|c|}
\hline
 & Task & Events & Resources & Event-pairs \\
\hline
\textit{Original log characteristics}  & 18 & 6870 & 561   & 1116776   \\
\textit{Multitasking log characteristics} & 17 & 1267 & 76    & 1039      \\ 
\hline
\% of characteristics with multitasking & 94.4\% & 18.4\% & 13.5\% & 0.09\% \\
\hline
\end{tabular}
\end{table}

As mentioned in Section~\ref{sec:formalization}, \textit{Multitasking Log Index (MTLI)} 
is another measure that helps identify the percentage of multitasking in the entire event log. 
For the \textit{ACR log}, $MTLI = 1.05\%$. If $MTLI$ is low, one would expect the rate of improvement in the analysis to be low as well, but as the level of multitasking in the log increases, the measure should improve proportionally. This indicates that when analyzing all possible pairs of events, only the 1.05\% of the time of those events were overlapped, which represents low level of multitasking. 
For the same log, the \textit{Multitasking Work Item Index} is $MTWII = 58.54\%$, that indicates that, for those events where multitasking has been identified, the pairs of events are overlapping by 58.54\% of their total duration. 

\begin{figure}[b]
  \begin{center}
    \includegraphics[width=\textwidth]{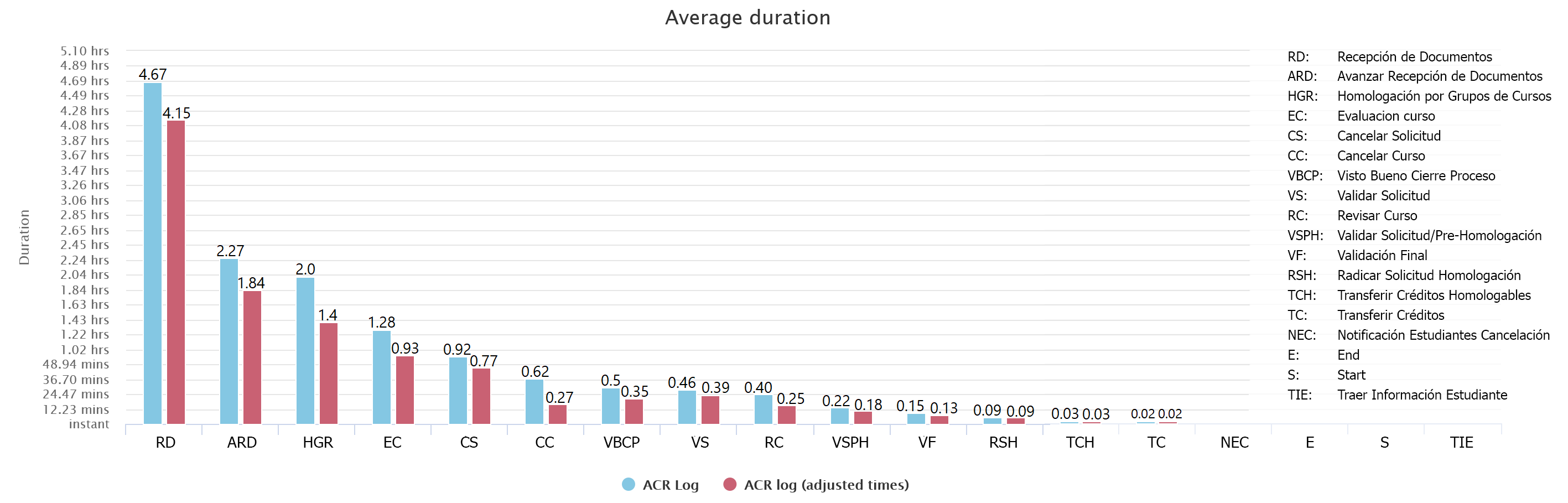}
    \caption{Comparison of average duration between \textit{ACR log} and \textit{ACR adjusted log}}
    \label{fig:chart_bothLogs_AverageDurations_Percentage}
  \end{center}
\end{figure}

Using Apromore, \textit{ACR log} and \textit{ACR adjusted log} were analyzed and compared in terms of time. 
Figure~\ref{fig:chart_bothLogs_AverageDurations_Percentage} shows the average duration of process tasks. Blue bars represent the average time duration of \textit{ACR log} tasks and red bars the average time duration of \textit{ACR adjusted log} tasks. 
Both \textit{RD} and \textit{HGR} reflect a difference of 0.6 hours (h), followed by \textit{HGR} with 0.43 h; \textit{EC} and \textit{CC} with 0.35h; \textit{CS}, \textit{VBPC} and \textit{RC} with 0.15 h; \textit{VS} 0.07; \textit{VSPH} 0.04; \textit{VF} with 0.02 h; \textit{RSH} does not show improvement; and the last 6 task do not reflect improvement either, but can be considered activities of instant duration. 
Finally, using BIMP, three resource pools were identified and slight differences in the percentage of resource utilization were noticed.

In general, the above results show that with the pre-processing of log it is possible to effectively 
reduce processing times of tasks and to maintain and/or improve similarity between traces involved in each log. Besides, we figured that the level of improvement in the results depends on the multitasking level in the log: the number of event pairs overlapped and the percentage of overlap between each event pair. 
As we have only been able to identify and use one real-life log with multitasking characteristics to show dependence between the amount of multitasking and the result improvements, in the following subsection, a synthetic log was modified to generate a log set with different multitasking~levels.


\subsection{Evaluation based on a Synthetic Log}
\label{subsec:evaluation_synthetic}

The objective of this section is to identify how the level of multitasking affects the discovery and simulation of BPS models.  
Our hypothesis is based on the assumption that the results of BPS models vary and are enhanced depending on the amount of overlap identified in each log. 
This scenario is composed of a set of event logs derived from a synthetic log called \textit{PurchasingExample.xes}. 
This is one of the public event logs available through the SIMOD distribution~\cite{Camargo_2019_SIMOD} that was generated from a purchase-to-pay process model not available to the authors. This event log, which does not contain multitasking characteristics (\textit{PE\_0P\_log}), has 608 traces, 21 tasks, 9119 events and involves 27 resources. 


\subsubsection{Experimental setup.}
\label{subsub:syntlog_setup}

The validation process if divided into the following steps:
\begin{enumerate}

\item \textit{Selection and preparation of the base log.} The \textit{PE\_0P\_log} base log does not contain multitasking characteristics. Therefore, when calculating their multitasking indexes they have a value of zero. 
New logs were generated using different percentage of shifting (overlap between events or \textit{work items}) for each log. 
To generate the new event logs, we implemented a Python script (\textit{percentage.py}) that, given a percentage of shifting (between 0.0 and 1.0) generates a new event log in XES format including events overlapped in that percentage of their processing times.
The script algorithm works as~follows. 

\begin{itemize}
\item The base event log is divided by grouping the events that are executed by a particular resource (see Description \ref{def:segment_per_resource}, \textit{segment per resource}). 
\item The events of each \textit{segment per resource} are ordered according to their start timestamps. 
\item For each \textit{segment per resource}, the first event is taken as the pivot and the next adjacent event is searched among the remaining events. Two events ($e_1, e_2$) are adjacent events if the end timestamp of $e_1$ has the same value as the start timestamp of $e_2$. In Figure \ref{fig:intersections_synthetic_log}.a, the first pair of events shown are adjacent events.
\item When a pair of adjacent events are identified, the timestamps are shifted depending on the percentage assigned.  
In Figure~\ref{fig:intersections_synthetic_log}.b, 20\% of shifting is applied, while in Figure~\ref{fig:intersections_synthetic_log}.c, the shifting is~40\%.

\begin{figure}[b]
  \begin{center}
    \includegraphics[width=\textwidth]{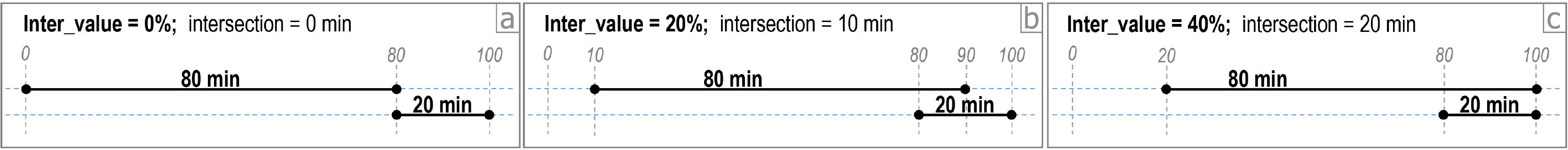}
    \caption{Overlapping of events according to a percentage of shifting.}
    \label{fig:intersections_synthetic_log}
  \end{center}
\end{figure}

\item The two events of a pair of adjacent events are excluded from the following search. The next event in the \textit{segment per resource} is taken as pivot and the search is repeated. If no adjacent event is found, it is not modified and the search is repeated with a new pivot.
\item The search of adjacent events is repeated for all \textit{segment per resources}.
\item The resulting log will have a multitasking percentage \textit{MTWII} similar to the percentage of shifting indicated in the script, although we will comment this value may vary. However, the total percentage of multitasking in the log (\textit{MTLI}), depends on the number of adjacent events identified.

\end{itemize}

\item \textit{Generate the set of adjusted logs.} The \textit{Sweeper} script is run on each of the logs generated in the previous step. 

\item \textit{Calculate measures.} 
\begin{itemize}
\item Using the \textit{Sweeper} script the multitasking indexes in the log were calculated (\textit{MTWII} and \textit{MTLI}). 
\item The hyper-parameter optimization of SIMOD was used with each log to obtain similarity measures in each case. To do that, 100 BPS models were generated using different setup combinations of processing  parameters. Parameters $epsilon$ and $eta$ varied from 0.0 to 1.0. Each simulation model was executed 5 times, for that, 500 simulations were evaluated for each log in the set.
\item BIMP can be used to analyze the resource utilization percentages.

\end{itemize}

\item \textit{Comparison and analysis of results.} 

\end{enumerate}


\subsubsection{Analysis of the results.}
\label{subsub:syntlog_analisis_results}

The set of synthetic event logs was made up of 6 logs. The \textit{PE\_0P\_log} and 5 more logs built using the \textit{Percentage} script with the \textit{PE\_0P\_log} as a base log and using a percentage of shifting of 5\%, 10\%, 15\%, 20\% and 25\%. 
With regard to resources, in \textit{PE\_0P\_log} participate 27 resources and 11 of them reflect multitasking. From the 9119 events in \textit{PE\_0P\_log}, 2625561 pairs of events were identified. From these, 789 are adjacent events to be used for time shifting of the logs, excluding in this set all instantaneous events. 

Table~\ref{tab:synthetic_log_comparison} shows the percentage of shifting applied to the adjacent events in each generated log; the multitasking indexes (\textit{MTWII} and \textit{MTLI}) for each log in the set; the number of pairs of events in which overlapping was identified (multitasking); and the value of two similarity measures obtained using SIMOD. \textit{DL-Mean Absolute Error} (\textit{DL-MAE}) assesses the similarity between two traces evaluating an attribute, in this case the \textit{cycle time} of traces, and the \textit{Mean Absolute Error} (\textit{MAE}) of the cycle time traces expressed in seconds.

The percentage reflected in the column \textit{MTWII} should be the same as \textit{Shifting}, because the number of adjacent events on which the shifts were made was the same for all the logs. However, certain \textit{MTWII} values are slightly higher since the shifting of some events may generate overlapping between events that initially were not adjacent. 
This is also reflected in the column \textit{Overlapping Pairs}, where the number of pairs of events overlapped is greater than 789 and increases as the shifting increases. 
Above a certain amount of shifting, the value of \textit{MTWII} is less than the percentage of shifting. This is because a shift can cause one event to be embedded within another (Figure \ref{fig:intersections_synthetic_log}.c), and if the shifting increases, the event is still embedded and does not provide more multitasking to the log. 
As could be deduced by identifying the number of events in the log, the \textit{MTLI} is quite low. However, like \textit{MTWII}, it increases when the percentages of shifting increase. 
Similarly, since the amount of multitasking in the log is low, the difference between \textit{DL\_MAE} values varies and improves slightly for those cases where the percentage of shifting is quite similar to \textit{MTWII} (0, 5, 10, 15) and worsen slightly for those cases where the shifting and index vary more. 
Finally, when calculating the MAE we see that although there is a significant difference between a log with and without multitasking, as the multitasking is increased, and the adjustment in the logs, the MAE is gradually reduced, which means that the discovered BPS models are more accuracy. 
BPS models were simulated using BIMP. 5 resource pools were discovered, two of them with high percentage of resource utilization (RU). The RU in BPS models derived from multitasking is reduced, especially for those resource pools where the RU is higher.

\begin{table}
\centering
\caption{Comparison between the synthetic logs created using a percentage of shifting.}\label{tab:synthetic_log_comparison} 
\begin{tabular}{|c|c|c|c|c|c|c|}
\hline
Shifting (\%) & MTWII (\%) & MTLI      & Overlapping Pairs  &  DL\_MAE & MAE (segs)\\
\hline
0         &   0       &   0       & 0         & 0.8883 & 1073208\\ 
5         & 5.596     & 1.468e-05 & 876       & 0.8889 & 1145181\\ 
10        & 10.381    & 2.754e-05 & 950       & 0.8893 & 1098788\\ 
15        & 14.694    & 3.953e-05 & 1006      & 0.8895 & 1091332\\ 
20        & 18.860    & 5.087e-05 & 1041      & 0.8841 & 1049593\\ 
25        & 22.266    & 6.147e-05 & 1073      & 0.8866 & 1117721\\ 
\hline
\end{tabular}
\vspace*{-5mm}
\end{table}

\section{Conclusion}
\label{sec:conclusions_futurework}

This paper outlined an approach to discover BPS models that take into account  multitasking behavior. Specifically the paper showed how to pre-process an event log in order to discover multitasking behavior and how to adjust the processing times of tasks in such a way that the resulting log does not contain multitasking behavior, yet the resource utilization in the resulting log is equivalent to that in the original log. In this way, the BPS model discovered from the pre-processed log takes into account the multitasking behavior but can be simulated using a traditional process simulator (e.g. BIMP). 

The evaluation showed that, in the presence of multitasking, the approach improves the accuracy of BPS models. 
We also identified that the greater the percentage of overlap in multitasking events in a log, the more the approach improves the accuracy of the generated BPS models. 
The experimental evaluation was restricted to one real-life and the amount of multitasking in this logs was low, so it was difficult to generalize the results. The evaluation on synthetic logs partially addressed this limitation by introducing varying levels of multitasking. Still, the obtained levels of multitasking remained relatively low due to the approach employed to add multitasking behavior in the synthetic log.

The discovery of simulation models is key to the setup of as-is scenarios that allow the reliable evaluation of what-if scenarios focused on process optimization. Processes are dynamic and the results of their execution may vary over time, largely due to the behavior of the human resources involved, and this characteristic is independent of the defined process model. Even in those cases in which the process is not clearly defined and only execution records are available, simulation models obtained allow the analysis of processing times or resource utilization rates, which can be influenced by human behavior such as multitasking, batching or delaying of low-priority tasks. 

A possible direction for future work is to extend the evaluation to other real-life logs with higher levels of multitasking. The challenge here is that event logs where both the start and end times of tasks are available are generally not available in the public domain. An alternative approach is to design new methods for generating realistic synthetic logs with high levels of multitasking.

The present work was limited to multitasking across multiple instances of one business process. Another avenue for future work is to discover and handle multitasking across multiple business processes. The latter would require the ability to simulate multiple business processes simultaneously.

\medskip\noindent\textbf{Acknowledgments.} This research is funded by the European Research Council (ERC Advanced Grant - Project PIX 834141) and the European  Commission (FEDER) and the Spanish R\&D\&I programmes (grants P12–TIC-1867 (COPAS), RTI2018-101204-B-C22 (OPHELIA))

%
%
%
\bibliographystyle{splncs04}
\bibliography{bibmultitasking}
\end{document}